\documentclass[pdflatex,sn-mathphys-num]{sn-jnl}

\usepackage{graphicx}%
\usepackage{multirow}%
\usepackage{amsmath,amssymb,amsfonts}%
\usepackage{amsthm}%
\usepackage{mathrsfs}%
\usepackage[title]{appendix}%
\usepackage{xcolor}%
\usepackage{textcomp}%
\usepackage{manyfoot}%
\usepackage{booktabs}%
\usepackage{algorithm}%
\usepackage{algorithmicx}%
\usepackage{algpseudocode}%
\usepackage{listings}%

\begin{document}
	
	\title[Article Title]{\centerline{Exoplanets synchronization in the habitable zone:}
		Learning from Venus' retrograde rotation}
	
	
	\author[1]{\fnm{S.} \sur{Ferraz-Mello}}\email{sylvio@iag.usp.br}
	
	
	\affil[1]{\orgdiv{Institute of Astronomy, Geophysics and Atmospheric Sciences, Universidade de S\~ao Paulo}, \city{S\~ao Paulo}, \postcode{05508-090}, \state{SP}, \country{Brasil}} 
	

\maketitle

\begin{abstract}
	
The rotation of a planet located in the habitable zone of a solar-type star can be reversed by a smooth process associated with the formation of its atmosphere and the increase of stronger torques, opposite to normal tidal torques.
Our understanding of the rotational dynamics of Venus is revisited to analyze what might happen to exoplanets in the habitable zone of a solar-type star.
The creep tide theory is used to calculate the gravitational tidal torque.
Mathematical analysis is used to study the differential equation resulting from the combined effects of tidal torque and atmospheric torque.
It shows that no collision with other bodies or critical planetary perturbations is necessary to convert the rotation of an Earth or super-Earth with a significant atmosphere formed during its evolution into a retrograde rotation.
The reversal of a planet's rotation is not an exceptional event and may have occurred many times among known exoplanets in the habitable zone.
It is sufficient for the planet to be at a sufficiently short distance from its host star to allow tidal torques to nearly synchronize the planet's rotation before most of its atmosphere forms (but not so close that stellar radiation destroys the formed atmosphere).
When atmospheric torques become more important than tidal torques, a pitchfork bifurcation occurs: the synchronous attractor bifurcates into two asynchronous attractors, and the system evolves toward one of the asynchronous attractors.
If it evolves toward the subsynchronous branch, the rotation may subsequently become retrograde. Venus's rotation is an example.
None of these processes is catastrophic.
Planetary atmosphere formation is a continuous and smooth process, which may be more or less efficient, but it is not a low-probability event.
\end{abstract}



\section{Introduction}

This paper aims to use our knowledge of Venus' rotation as an insight into what to expect for the rotation of Earth-like exoplanets in the habitable zone of a solar-type star. It collects various published results and compares them to hypothetical situations to improve our understanding of their meanings. It includes a new discussion of the existence and stability of stationary solutions under the joint actions of tidal and atmospheric torques. When the eccentricity is neglected, these solutions might be synchronous or asynchronous (a.k.a. parasynchronous), stable or unstable (Correia and Laskar, 2001, 2003a, 2003b; Correia et al. 2003; Leconte et al. 2015, Auclair-Desrotour et al. 2017). Two parameters indicating the ratios of tidal and atmospheric torques are introduced, along with a toy model. They allow us to study several different evolutionary paths for the rotation of exoplanets located in the habitable zone of solar-type stars. In these cases, the HZ is not too close to the star, and planets may harbor dynamically important atmospheres. The exploration of the model shows that the evolution leading to a retrograde rotation can happen at a slow pace and is not a low-probability event. The retrograde rotation of Venus may be the result of a smooth, deterministic process associated with the formation of its atmosphere, rather than a collisional process that suddenly transferred a huge amount of negative angular momentum to the planet. The collisional processes often used to explain the retrograde rotation of Venus (McCord, 1968; Makarov and Goldin, 2023) are not the only possibilities. The smooth process proposed in this paper was first described by Correia and Laskar (2003b).
It is not the only way to reverse the rotation of a planet, but it is simple, realistic, and does not depend on exceptional conditions, such as high eccentricity or obliquity.

In section 2, we present the main data on Venus' rotation, and the new full 3D version of the creep tide theory (Folonier et al. 2025) is used to determine how long the retrograde rotation can resist the action of the tidal torque. In section 3, the torque due to the phase lead of the atmospheric bulge is introduced, and the main effects due to the joint actions of the atmospheric and tidal torques are investigated. In section 4, it is shown how the evolution of the atmosphere can cause the reversal of the planet's rotation from direct to retrograde. A toy model for the planet rotation provides evolutionary examples of this process. Section 5 collects the main conclusions of the report. An appendix shows the that the use of the classical Darwin's tidal theory to calculate the tidal torque, leads to the same results as the creep tide theory. 

\section{Venus rotation}\label{sec:Naked}
Venus currently has a retrograde rotation with a period of $243.0226 \pm 0.0013$ days  (Margot et al. 2021). We know that this rotation is strongly affected by the thermal bulges in the planet's atmosphere. How would a retrograde rotation of this kind evolve should the planet be devoid of an atmosphere? We have investigated this topic using the new version of the creep tide theory valid for systems with a nonzero obliquity (Folonier et al. 2025).

\subsection{Back to prograde}
\begin{figure}[t!]
\centering
\includegraphics[width=0.7\textwidth]{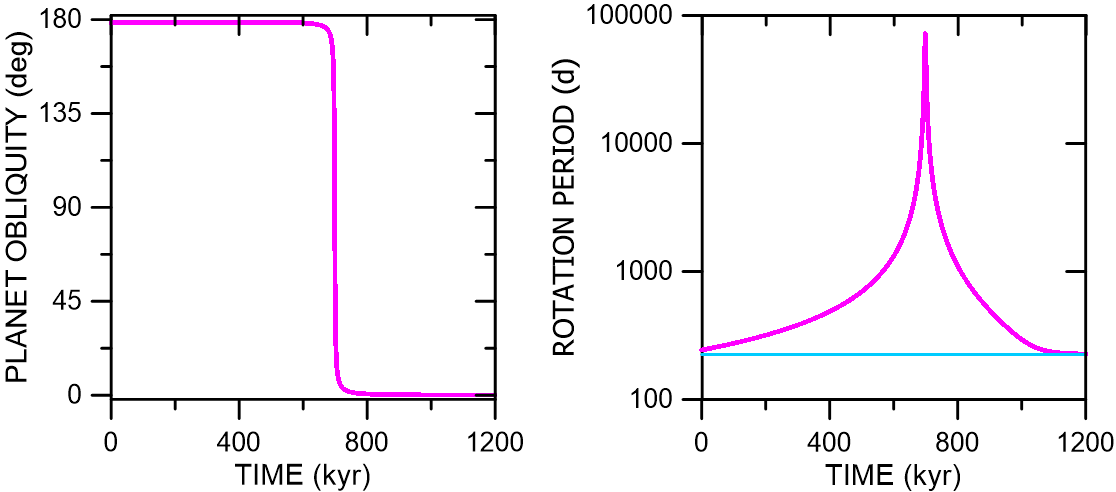}
\caption{Rotation Reversal: {\it Left:} Tidal reversal of the planet obliquity; {\it Right}: Evolution of the rotation period (red) and the orbital period (blue). } \label{fig:Nova}
\end{figure}

The main result is that the rotation of a naked Venus would be changed to prograde in a relatively short time (see Fig. \ref{fig:Nova}). Using typical values for Venus' parameters involved in this calculation (see Table \ref{tab:1}), the result is an inversion in $\sim 700,000$ years, if the relaxation is $\gamma \sim 2\times 10^{-7} {\rm s}^{-1}$ (as the solid Earth {\textit cf.} Ferraz-Mello, 2013). The tidal theory also indicates that at the inversion, we do not have a reversal of the rotation but a continuous migration of the rotation poles, which are displaced until the inversion is completed. During the inversion interval, the angular momentum of the planet is very small (the rotation period grows to several thousand days). We do not know if the planet itself would turn upside down because the equations of the polhode are not included in the model.
After the reversal of the rotation back to a prograde state, the rotation accelerates and is driven to a stable synchronization with the orbital period (see Fig. \ref{fig:Nova} {\it Right}).

\begin{table}[t]
\caption{Venus Dynamic Parameters} \label{tab:1}
\begin{tabular}{@{}ll@{}}
\toprule
Mass ($M$)& 0.815 Earth mass\\
Radius ($R$) & 0.949 Earth Radius \\
Rotation Period & 243.0226 d  (R)\\
Oblateness & $6.64 \times 10^{-6}$\\ 
Axial tilt with respect to the ecliptic pole ($J$) &  $175^\circ .7$\\
Obliquity & 178$^\circ.3933$ \\
Moment of Inertia ($C$) & 0.34 $MR^2$ \\
Tidal Relaxation Factor ($\gamma$) & $2 \times 10^{-7}\  {\rm s}^{-1}$ \\
Tidal Love number ($k_2$) &  0.295  \\   
Semi-major axis & 0.7238  AU \\
Orbital period & 224.9145 d\\
Eccentricity & 0.006773 \\
Inclination & $3^\circ.395$  \\
Length of the day & 116.8091 d \\
Atmospheric Bulge Lead ($\sigma'$) & $30-34^\circ$ \\
Semi-diurnal frequency ($\nu)$ & $-1.245144 \times 10^{-6} {\rm s}^{-1}$ \\
\botrule
\end{tabular}  
\end{table}

\subsection{The rotation phase diagram}

In the zero-obliquity version (a.k.a. coplanar version) of the tidal theory, the average variation of the rotation of the planet is given by the first-order differential equation (Ferraz-Mello, 2013; Correia et al. 2014; Ferraz-Mello et al. 2022)  
\begin{equation}
\langle\dot{\Omega}\rangle = -A \sum_{k \in Z} E^2_{2,k} \sin 2\sigma_k
\label{eq:tidal}
\end{equation}
where 
\begin{equation}
 A = \frac{3k_2GM^2R^5}{4Ca^6}
 \label{eq:A}\end{equation}
is a positive coefficient depending on the dynamical parameters of the system: $G$ is the gravitational constant, $M$ the mass of the central star, $R$ the planet radius, $C$ its moment of inertia, $k_2$ the tidal Love number, and $a$ the orbital semi-major axis. $E_{2, k}$ are known functions of the orbital eccentricity $e$ (Cayley polynomials), and $\sigma_k$ are the angular lags of the components of the tide, defined in the interval $(-\pi/2,+\pi/2)$, by the functions
\begin{equation}
\tan \sigma_k = \frac{\nu + kn}{\gamma}
 \label{eq:lag}
\end{equation}
where $\nu=2\Omega-2n$ is the semi-diurnal frequency.
Hence,
\begin{equation}
\sin 2\sigma_k = \frac{2\gamma(\nu + kn)}{\gamma^2+ (\nu + kn)^2}.
\end{equation}
In the circular approximation, the only remaining term is $k=0$ and $\sigma_0$ is the lag of the main tidal component: the semi-diurnal tide.  

\begin{figure}[t!]
\centering
\includegraphics[width=0.6\textwidth]{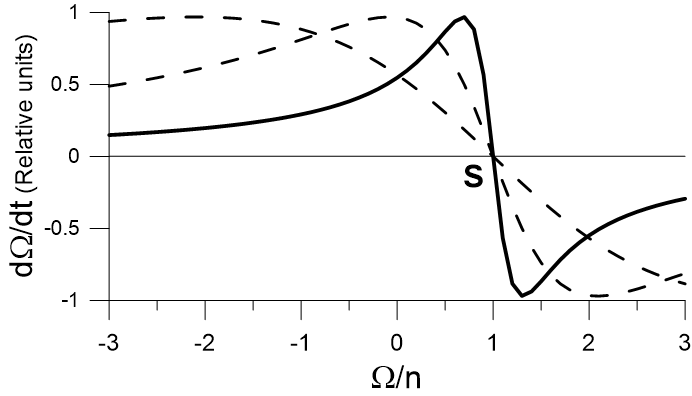}
\caption{Phase diagram for $e=0.006773$ and relaxation factors $2\times 10^{-7} {\rm s}^{-1}$ (solid line), $7\times 10^{-7}{\rm s}^{-1}$ and $2\times 10^{-6}{\rm s}^{-1}$ (dashed lines).} \label{fig:omega1}
\end{figure}

Figure \ref{fig:omega1} is the plot of the function $\langle\dot{\Omega}\rangle = f(\Omega)$ for a wide range of rotation velocities of a naked Venus.
It shows that for the current orbital eccentricity, there is only one stationary solution $\mathsf{S}$, at $\Omega/n \simeq 1$. This solution is stable since $\dot{\Omega}$
is positive on the left of $\mathsf{S}$ and negative on its right. 
This situation is very robust and is verified for a wide range of variations of the parameters involved in the function $f(\Omega)$. 

\section{The atmospheric torque}\label{sec:atmosf}
The retrograde rotation of Venus could not persist for a long time if the gravitational tidal torque was the only one acting on the planet. The observed retrograde rotation is a clear indication of the existence of a torque acting in the opposite direction. It has been attributed to a phase lead in the thermal deformation of the atmosphere (Gold and Soter, 1971; Ingersoll and Dobrovolskis, 1978).\footnote{For detailed discussions of this model, see  Correia and Laskar, 2003a, Leconte et al. 2015, Auclair-Desrotour et al. 2017, 2024, and references therein.}  

The solar attraction also produces deformations in the atmosphere of Venus, giving origin to an additional torque braking the planet's rotation (Dobrovolskis, 1980; Valente and Correia, 2023).
However, in the absence of a significant obliquity, it is weaker and not considered in the present study.

\begin{figure}[t]
	\centering
	\includegraphics[width=0.6\textwidth]{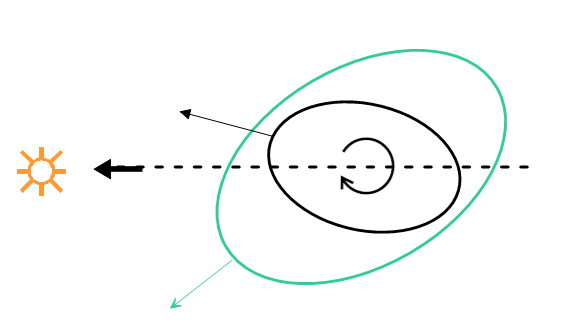}
	\caption{The atmospheric and tidal bulges in a system of reference in which the direction of the Sun is kept fixed. \textit{cf.}  Ingersoll and Dobrovolskis, 1978.} \label{fig:bulges}
\end{figure}

Solar radiation raises the air temperatures in the afternoon and causes the air mass to flow to colder regions.  In a first approximation, the atmosphere may be represented by an ellipsoid whose bulge is displaced in the direction contrary to the rotation of the body 
(see fig. \ref{fig:bulges}). In this figure, the direction of the orbital motion does not matter. The rotation direction indicated is not absolute, but relative to the Sun's direction. At one given point on the planet's surface, the tidal bulge is felt after noon. The gravitational tide is delayed. In contrast, the atmospheric bulge leads the rotation.
The gravitational attraction of the Sun on the atmosphere introduces a torque in a direction opposite to that of the gravitational tide. 

Using notations similar to those adopted in the previous Section, the contribution of the atmospheric bulge to the rotation of the planet is  
\begin{equation}
	\dot{\Omega} = A' \sin 2\sigma'
\label{eq:atm}\end{equation}
where 
	\begin{equation}
		A'\gamma'=\frac{3\pi n^2 R^4 \Phi_0}{8c_pT_0 C}. 
	\end{equation}
is a positive coefficient depending on the mean motion ($n$), the radius ($R$), the moment of inertia ($C$), on the temperature at the surface,  ($T_0$), on its specific heat at constant pressure, ($c_p$), and on the heat flux at the sub-solar point, ($\Phi_0$).\footnote{This coefficient formally differs from the coefficient appearing in   Ingersoll and Dobrovolskis (1978) by a factor 2. The division by 2 is here necessary to compensate for the factor 2 present in the expression of $\sin 2\sigma'$.}
$\sigma'$ is the lead of the atmospheric bulge defined by
\begin{equation}
	\tan \sigma' = \nu\tau' \equiv \frac{\nu}{\gamma'}
\label{eq:lead}\end{equation}
 (Ingersoll and Dobrovolskis, 1978) where $\tau'$ is a proportionality constant and $\gamma'=1/\tau'$. 
$A'$ has the same dimension as $A$. Note that when $e=0$,  equations (\ref{eq:tidal}) and (\ref{eq:atm}) are formally equal. As before, the angle $\sigma'$ is defined by the principal value of the function, in the interval $(-\pi/2,+\pi/2)$.
 In the case of Venus, the maxima of the atmospheric pressure are recorded at 09:45 and 22:00 local solar time (see  Auclair-Desrotour, 2024). This means $\sigma' \sim 32^\circ$ . 

\subsection{Composite phase diagram}\label{sec:composite}

Under hypotheses of zero obliquity and zero eccentricity, the variation of the rotation of the planet is given by an autonomous first-order differential equation formed by the composition of Eqns. (\ref{eq:tidal}) (for $e=0$) and (\ref{eq:atm}):
\begin{equation}
	\dot{\Omega} = -A \sin 2\sigma_0 + A' \sin 2\sigma'
	\label{eq:compos}
\end{equation}
or
\begin{equation}
	\dot{\Omega} = A \sin 2\sigma_0 \left(b\ \frac{\nu^2+\gamma^2}{\nu^2+\gamma'^2} -1 \right)
	\label{eq:nova}\end{equation}
where 
\begin{equation}
	b=\frac{A'\gamma'}{A\gamma}.
	\label{eq:b}\end{equation}
The parameter $b$ is akin to the ratio of the two torques. It is $b=0$ in the absence of atmospheric torque, and it increases with the influence of the atmospheric torque. It plays a critical role in the existence and stability of stationary solutions.

\begin{figure}[t]
	\centering
	\includegraphics[width=0.5\textwidth]{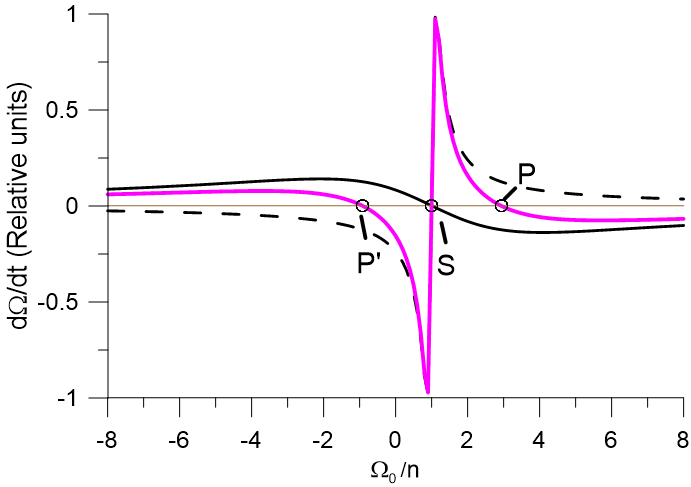}
	\caption{ Phase diagram $\dot{\Omega}(\Omega)$ resulting from the composition (red) of the tidal (solid black) and atmospheric (dashed) torques. The intersections of the red curve with the axis $\dot{\Omega}=0$ (brown) are the stationary solutions: The unstable synchronous solution $\mathsf{S}$ and two stable asynchronous solutions $\mathsf{P}$ and $\mathsf{P}'$, one prograde and one retrograde. The adopted factors $\gamma$ and $\gamma'$ were exaggerated to improve the visibility of the plot's main features.} \label{fig:composit}
\end{figure}

The red curve in Fig. \ref{fig:composit} is the plot of the resulting function $\dot{\Omega}(\Omega)$ in arbitrary units (such that the largest elongation of the function is equal to 1). The figure also shows the individual contribution of the two terms of Eqn. (\ref{eq:compos}): the tidal torque (black solid-line) and the atmospheric torque (dashed line). 

The intersections of the red curve with the axis $\dot{\Omega}=0$ are stationary solutions of the system. One of these solutions, $\mathsf{S}$, lies at $\Omega/n = 1$ and is a synchronous solution. A simple analysis of the sign of $\dot\Omega$ on the right and left of this point indicates that this solution is unstable. The derivative of the curve $\dot{\Omega}(\Omega)$ in the neighborhood of the crossing point is positive. One may note that in the component corresponding to the tide of a naked planet (black solid line), the sign is negative, indicating the stability of the synchronous solution in that case. 

The figure shows two more intersections at the points $\mathsf{P}$ and $\mathsf{P}'$. The derivative of the functions at these two crossings of the axis $\dot{\Omega}=0$ is negative, so both correspond to stable stationary solutions. These two asynchronous solutions are symmetric with respect to the synchronous solution and, in the case shown in Fig. \ref{fig:composit}, the solution at $\mathsf{P}$ is prograde and the solution at $\mathsf{P}'$ is retrograde.

In the case of Venus, the figure is similar to Fig. \ref{fig:composit}. However, the excursion of the function $\dot{\Omega}(\Omega)$ near the synchronous solution is too large and, keeping the figure at scale, the red lines near the asymmetric solutions and the axis $\dot{\Omega}=0$ are indistinguishable, leaving the crossings invisible (even if they are well defined in the data file). To make them visible, the factors $\gamma$ and $\gamma'$ adopted in the construction of Fig. \ref{fig:composit} were multiplied by 10.

\subsection{Existence and stability of the asynchronous solutions}\label{sec:async}

The bracket in Eqn. (\ref{eq:nova}) can be expressed as the quotient of two polynomials in $\nu$, and equation $\dot{\Omega}=0$ has the same roots as the symmetric third-degree polynomial in the numerator. One of these roots is $\nu_0=0$, and the other two are given by
\begin{equation}
	\nu_0^2=\gamma'^2\ \frac{1-b\Gamma^2}{b-1}
\end{equation}
where
\begin{equation}
	\Gamma=\frac{\gamma}{\gamma'}.    
\end{equation}

The condition for the existence of real roots  $\pm\nu_0$ is that the numerator and denominator are simultaneously positive or negative (otherwise, we get $\nu_0^2<0$ and the two roots are imaginary):  
\begin{description}
	
	\item[Case I] $\ \qquad \Gamma^{-2}<b<1 \qquad (\Gamma>1) $;  
	\item[Case II] $\qquad 1<b<\Gamma^{-2}  \qquad(\Gamma<1) $.  
\end{description}
Otherwise, the two roots are imaginary.

The analysis of the behavior of $\dot{\Omega}(\Omega)$ when $|\Omega|\rightarrow\infty$ allows us to conclude that, in case I, the synchronous root is unstable and the two asynchronous roots are stable (as shown in Fig. \ref{fig:composit}). In case II, the behavior is inverted, the synchronous root is stable, and the two asynchronous roots are unstable. 

\begin{figure}[t]
	\centering
	\includegraphics[width=0.5\textwidth]{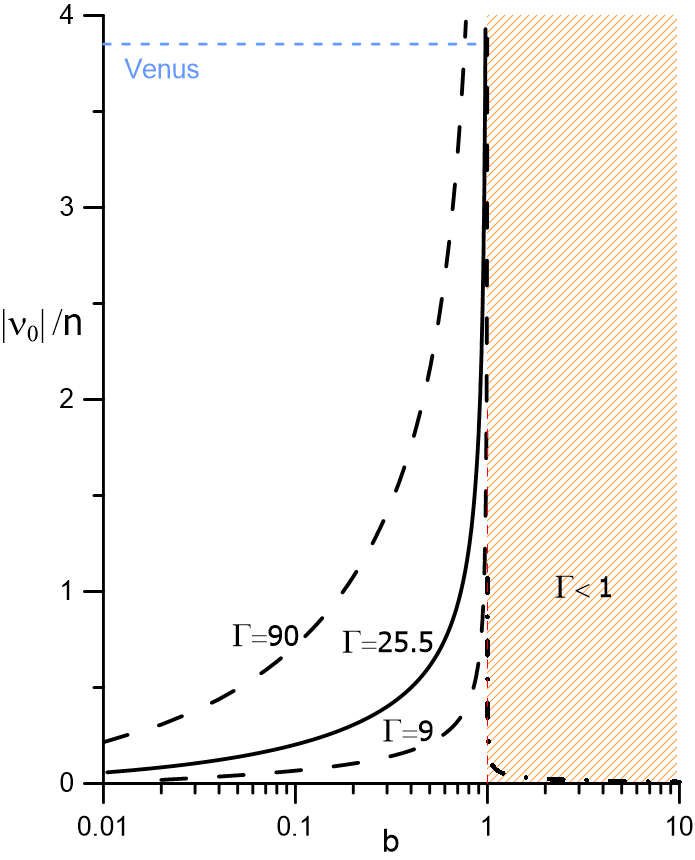}
	\caption{Loci of the asynchronous stationary solutions for given values of $\Gamma$. These solutions are stable if they fall in the zone $b<1$ and unstable otherwise.
		If $|\nu_0/n| < 1$, they are both prograde. If $|\nu_0/n| > 1$, one is prograde and the other is retrograde.} 
	\label{fig:parasync}
\end{figure}

\begin{figure}[t]
	\centering
	\includegraphics[width=0.5\textwidth]{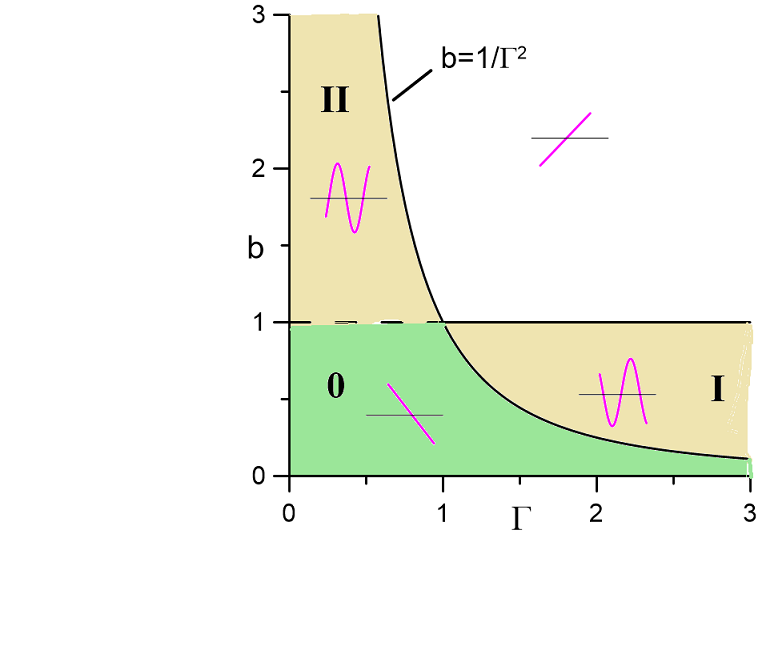}
	\caption{ Domains of existence of the asynchronous stationary solutions: I stable, II unstable. Outside these domains, they are not real. In domains 0 and II, the synchronous stationary solution is stable. In other domains, it is unstable. The inserted sketches show how the roots are formed in each domain. } \label{fig:dominios}
\end{figure}

These stationary solutions are similar to those found by Auclair-Desrotour et al. (2017) using a Maxwell rheology. 

\subsection{Stability of the synchronous solutions}

In domains I and II the synchronous solution is unstable when the asynchronous solutions are stable, and vice versa. In the other domains, the synchronous solution is the only real root. The stability is again deduced from the study of the leading term of the third-degree polynomial that defines the roots: $\dot{\Omega} = {\rm const}\cdot
(b-1)\nu^3 + \cdots $\footnote{The function is of the kind y'=y.f(y). The stability of the root $y=0$ may be inferred from the study of a finite interval.} The synchronous solution is stable when $b<1$ and unstable otherwise.
This is consistent with the plots of the partial contributions of the tidal and atmospheric torques shown in Fig. \ref{fig:composit}. 

In the domain 0 (green area) of Fig. \ref{fig:dominios}, the tidal torque is dominant and the synchronous solution is stable as in the case of a naked Venus. In the white area, the atmospheric torque is dominant, and the synchronous solution is unstable.

\section{The evolution towards a retrograde rotation}

In the general case, planets are assumed to have a prograde rotation (see Dones and Tremaine, 1993; Kokubo and Ida, 2007). 
Figure \ref{fig:composit} shows that in case I, it is impossible for the rotation to evolve smoothly to a retrograde condition. The space between the parasynchronous solutions $\mathsf{P}$  and $\mathsf{P}'$ acts as a wall that prevents one solution starting on the right-hand side (prograde) from reaching the left-hand side of the diagram.
The same situation is repeated more or less in the other cases. 
For this reason, we often read that some collisional event may be at the origin of Venus' retrograde rotation, the most realistic being related to the transfer of negative angular momentum from a retrograde satellite (see e.g. McCord, 1968; Makarov and Goldin, 2023).

However, the impossibility mentioned above results from considering a frozen model. 
The inspection of Fig. \ref{fig:dominios} shows that, if we accept that the physical conditions are certainly not frozen, and the relative contribution of the tidal and atmospheric torques may have been different in the past, it is possible to have paths leading smoothly from a  prograde rotation in the past to an asynchronous retrograde rotation in the present. We may, for example, consider an evolutive path starting in domain 0 (green area) of Fig.  \ref{fig:dominios} and crossing the frontier between it and domain I. In domain 0, the rotation will evolve towards the synchronous solution and will reach it if enough time is allowed for the path segment inside domain 0. When crossing the boundary towards domain I, the synchronous solution becomes unstable, and the rotation is forced to evolve towards one of the two new stable asynchronous solutions. If the rotation is synchronous before the crossing, there is no mathematical reason to privilege one of the two solutions, and we have equal probabilities of evolving towards $\mathsf{P}$ or $\mathsf{P}'$. However, the increase in the moment of inertia due to the transfer of mass from the planet's core to the atmosphere may create a bias in favor of the branch leading to retrograde rotations.

If the atmosphere of Venus is still changing, the current rotational state of Venus can change. For instance, if we assume that Venus' rotation is currently trapped in a retrograde asynchronous stationary solution, we get $b=0.97489$, which is very close to the boundary $b=1$. A small change, say, in the surface temperature of Venus, may be such that $b$ increases to $b>1$. Hence, stable asynchronous stationary solutions cease to exist, and the rotation of Venus will grow continuously.

In fact, we cannot a priori exclude the possibility that the rotation of Venus is already in such a state of runaway growth. The measurements of Venus' rotation period are difficult, and the available ones (see Margot et al. 2021) do not allow for the exclusion of a very slow growth.

\subsection{Smooth transitions in a toy model}\label{sec:toy}

We rewrite Eqn. (\ref{eq:nova}) with the modified dependent variable 
\begin{equation}
	y=\frac{\nu}{\gamma'}
	\label{eq:y}
\end{equation}
and the positive coefficient 
\begin{equation}
	D=\frac{4A\Gamma}{\gamma'}.
	\label{eq:D}
\end{equation}
Hence,
\begin{equation}
	\dot{y}=D\ \frac{(b-1)y^3+(b\Gamma^2-1)y}
	{(y^2+1)(y^2+\Gamma^2)} \label{eq:doty}
\end{equation}

\begin{figure}[t]
	\centering
	\includegraphics[width=0.7\textwidth]{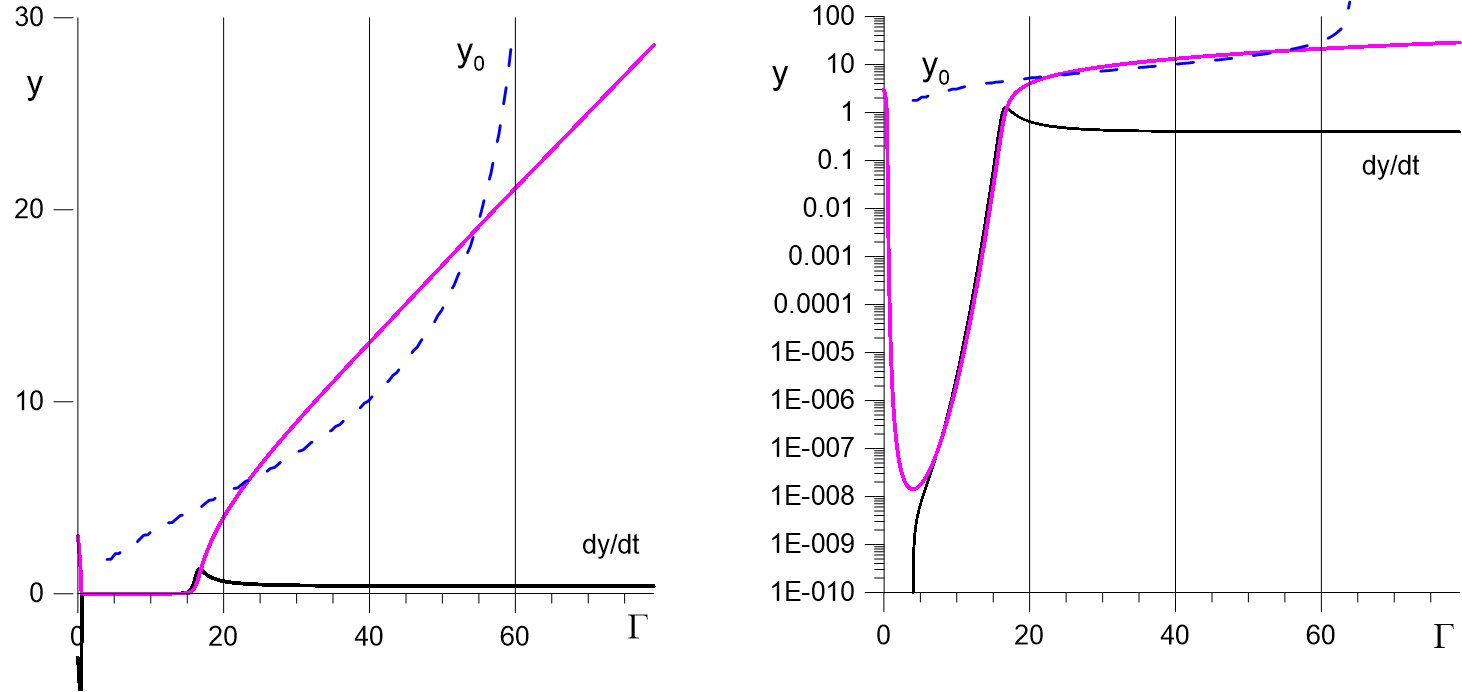}
	\caption{{\it: Left} Solution of Eqn. (\ref{eq:doty}) along a path defined by $b=\Gamma/64$ crossing the regions 0 and I of Fig. \ref{fig:dominios} before reaching the boundary $b=1$. {\it Right} The same is shown with a logarithmic scale. Also shown are the locus of the stationary root $y_0=\nu_0/\gamma'$ (dashed line), and the derivative $dy/dt$ (black). }
	\label{fig:toy}
\end{figure}

Fig. \ref{fig:toy} shows a solution for this equation along a path defined by $b=\Gamma/64$ with initial conditions $y(0)=3$ and $D=10$. This path is arbitrary, starts at the origin $b=0, \Gamma=0$ (absence of atmosphere), and evolves linearly as the atmosphere is formed. It crosses the boundary of domains 0 and I when $\Gamma=4$ and reaches the domain $b \ge 1$ when $\Gamma=64$. The following features may be seen: While $\Gamma < 4$,  $y$ decreases tending to the synchronous solution $y=0$, which is not reached; When $\Gamma = 4$, the solution $y=0$ becomes unstable, and $y$ evolves away from it. In the first segment, the growth is exponential (we have $\dot{y}\simeq y$). It approaches the stationary solution $\mathsf{P}$. Since the parameters $b$ and $\Gamma$ are continuously changing, the position of this solution changes (see the dashed line). It is not reached exactly. The actual solution oscillates around it. Nothing is seen when the boundary $b=1$ is crossed. The rotation velocity continues to grow at the same pace as before.

If the initial condition is taken as negative, the result is the same upside down. Equation (\ref{eq:doty})
can be written as $\dot{y}=yF(y,t)$, and it is easy to see that no solution can cross the line $y=0$. 
If at a given moment $y=0$, it will remain equal to 0 forever.

The toy model can be improved by taking into account that the atmosphere is formed by the planet's outgassing. So, mass is being transferred from the planet's inner parts to the atmosphere. This process increases the system's moment of inertia. The effect of a small uniform increase in the planet's moment of inertia is a small subtracting quantity $-Z $ in $dy/dt$. Hence, 
\begin{equation}
	\dot{y}=D\ \frac{(b-1)y^3+(b\Gamma^2-1)y}
	{(y^2+1)(y^2+\Gamma^2)} - Z 
	\label{eq:dotyZ}.
\end{equation}
where $Z$ is proportional to the rate of deceleration of the planet's rotation.

\begin{figure}[t]
	\centering
	\includegraphics[width=0.4\textwidth]{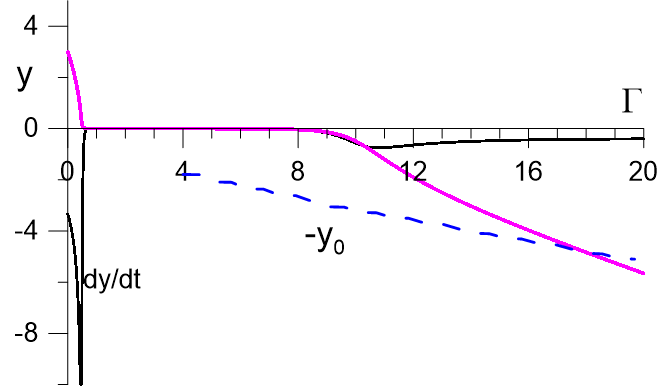}
	\includegraphics[width=0.4\textwidth]{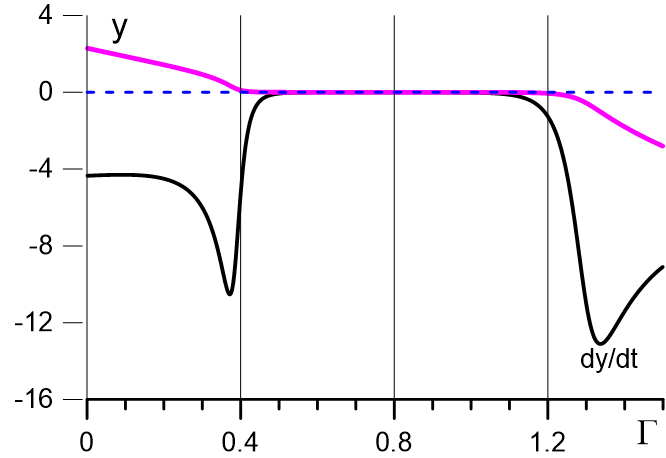}	\caption{Solution of Eqn. (\ref{eq:dotyZ}).
	At $t=0\ (\Gamma=0)$, the rotation is prograde. Near synchronization, it changes from supersynchronous ($y>0$) to subsynchronous ($y<0$). Subsequently, it becomes retrograde. {\it Left:} The same path as Fig. \ref{fig:toy}. {\it Right:}  Along a path that drives the solution from domain 0 to domain II, instead of domain I, and eventually to the white domain where the synchronous rotation is unstable. In both cases,  $Z=-10^{-3}$. }
	\label{fig:cross}
\end{figure}

Fig. \ref{fig:cross} {\it Left} shows the initial part of one solution for this equation obtained with the same parameters and initial conditions as before for $Z=10^{-3}$. In this solution, $y$ becomes negative as soon as the solution approaches the synchronous stationary solution, still inside the domain 0. The rotation becomes sub-synchronous (i.e. $\nu<0$, or $\Omega<n$) and, later on, becomes retrograde. We stress the fact that what is being observed in Fig. \ref{fig:cross}  is not the transition from prograde to retrograde, but the transition from super-synchronous ($\Omega > n$) to sub-synchronous ($\Omega < n$). But as $|\nu|$ increases, it can become larger than $2n$ and, hence,  $\Omega<0$.  

Fig. \ref{fig:cross} {\it Right} shows an alternative path, in which $b$ grows more quickly than $\Gamma$ ($b=2.5\Gamma$). While in the previous case, the increase in the angle $\sigma'$ led the evolution, here that angle changes slowly, and the evolution is ruled by the increasing coefficient $A'$, which means stronger atmospheric tides. The path goes from domain 0 to domain II (instead of domain I), but later on it crosses the boundary $b=\Gamma^{-2}$ and enters the white region where no stable stationary solution exists. $y$ remains near the synchronous solution for a while, but eventually starts a runaway growth to negative values.
 The rotation becomes sub-synchronous (i.e. $\nu<0$ , or $\Omega<n$) and evolves towards a retrograde one. 

\section{Conclusion}

The main result of this paper is that no collision with other bodies is necessary to convert the rotation of an Earth or super-Earth with an important atmosphere formed in the course of its evolution into a retrograde rotation. It is enough for the planet to be at a distance from the host star short enough to allow the tidal torques to almost synchronize the rotation of the planet before the formation of the bulk of its atmosphere (but not so close that the stellar radiation destroys the formed atmosphere). As the atmosphere is formed from the outgassing of the planet, the transfer of mass from the planet to its atmosphere may be responsible for a loss of angular velocity, so that the rotation is sub-synchronous when the atmospheric torques become more important than the tidal torques \footnote{Keep in mind that sub-synchronous means $\nu<0$ and retrograde means $\Omega<0$}. In the continuation, the prevailing atmospheric torques increase the relative rotation of the planet, allowing the rotation to become retrograde later on. None of these processes involves a catastrophic event (even if the transformation of a stable stationary solution into an unstable one is a typical example of René Thom's catastrophes -- see  Arnold, 1984). The formation of the planetary atmosphere is a continuous, smooth process, which may be more or less efficient but is not a low-probability event.

The following scheme (not included in the published paper) illustrates the pitchfork bifurcation.

\bigskip

\hspace{2cm}	\includegraphics[width=0.6 \textwidth]{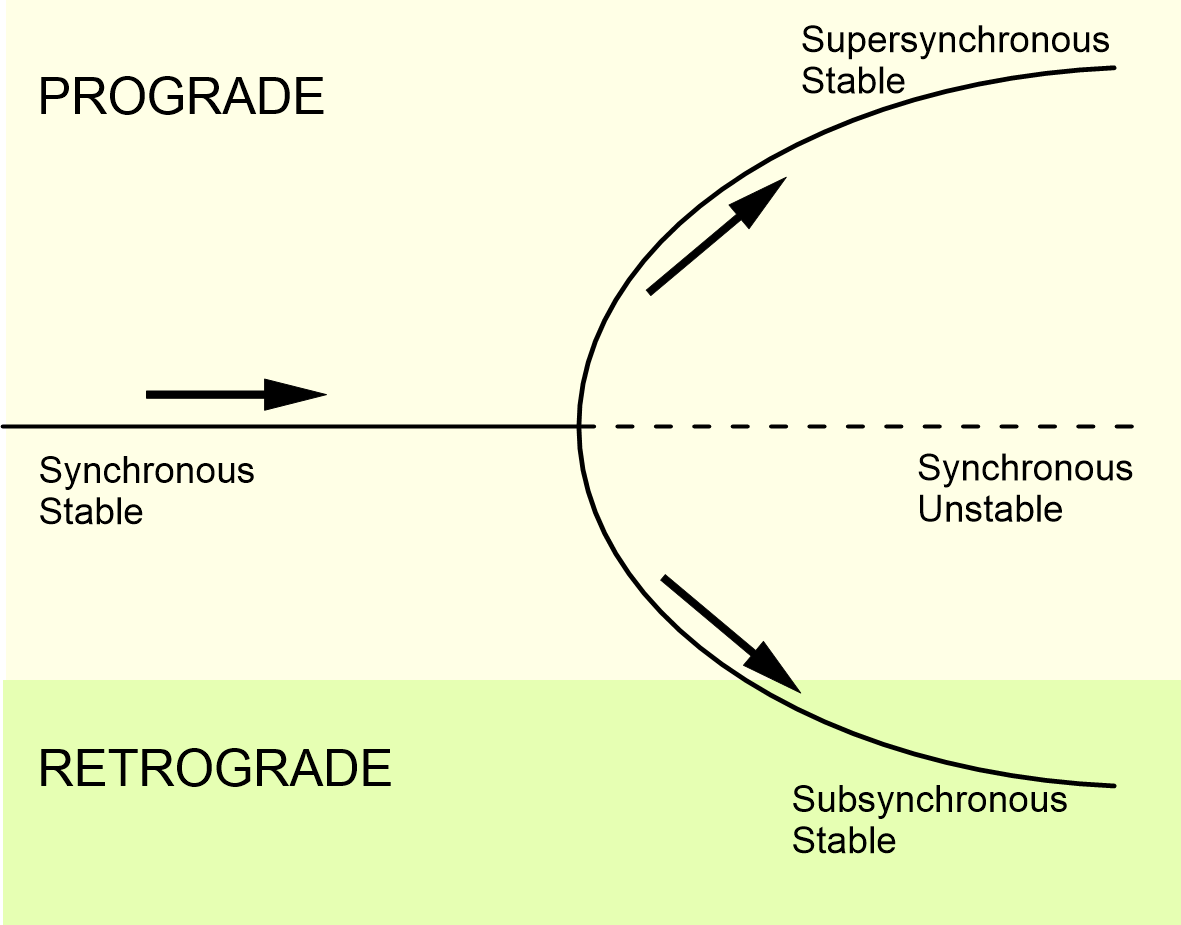}
\bigskip
\medskip

The examples constructed using a toy model in Sec. \ref{sec:toy} introduced evolution histories of the planet in which the parameters $b$ and $\Gamma$ grow uniformly. This choice is only justified by its simplicity. The mathematical phenomena shown in that example do not depend on the adopted path. The essential points are that the system must remain in domain 0 for a long time so that it reaches an almost synchronization and then crosses the boundary $b=1/\Gamma^2$, where the nature of the synchronous solution changes from stable to unstable. The shift to retrograde is not a necessary fate. 
It depends on the sign of $\nu$ just before crossing. If the motion is still super-synchronous, it will remain prograde forever. Conversely, if it is sub-synchronous, it may evolve towards a retrograde rotation.

The conclusions come from simplified models, but they are very robust and should not be changed if different tidal models are used (see Correia and Laskar, 2003b; Correia et al. 2003). The gravitational tidal torque creates one stationary synchronous solution at $\nu \simeq 0$ as long as the eccentricity of the orbit is kept small. The existence of a contrary torque maintains the current retrograde rotation of Venus. Improvements in the atmospheric torque model should not change the fact that, acting alone, it creates an unstable synchronous solution. When the two torques act together, they lead to a synchronous solution that is stable as long as the gravitational tidal torque dominates and unstable otherwise. The result is a mere consequence of these model-independent facts.

The toy model introduced in Section \ref{sec:composite} owes its simplicity to the zero eccentricity and to the co-planarity of the planet’s orbit and equator. If we consider orbits with significant eccentricity, even if we keep the co-planarity hypothesis, many other equilibrium solutions appear. 
In the case of a planet without atmosphere, they are the well-known spin-orbit resonances $\Omega/n = k/2\quad (k=1,3,4,\cdots)$, the number of resonances increasing for larger eccentricities (see Correia et al. 2014, Ferraz-Mello, 2015). If, in addition, we have a non-zero obliquity, the scenario becomes yet more complex. It is then difficult to construct a general model, and it is preferable to consider particular cases, as the extended study of the stationary solutions of Kepler 1229 b done by Valente et al. (2924).

The dense atmosphere of the planet plays a role not only in the inversion of the planet's rotation but also in keeping it retrograde. The study of a naked Venus shows that the current retrograde rotation would become prograde again in the absence of the atmospheric torque in less than one million years. The full 3D creep tide theory indicates that, at this inversion, we do not have a simple reversal of the rotation but a continuous migration of the rotation poles, which are displaced until the inversion is completed. 

\bmhead{Acknowledgements}
This investigation was sponsored by CNPq (Proc. 303540/2020-6) and FAPESP (Proc. 2016/13750-6 ref. PLATO mission).  Part of this research was conducted at the Southwest Research Institute in Boulder. I thank R. Alves-Silva and the referee for their suggestions and comments.

\appendix

\section{Comparison with results obtained with Darwin's theory}

The use of some different tidal theories led Leconte et al. (2015) to results indicating that the synchronous solutions are always stable. This is a crucial question since, in such cases, it is impossible to obtain results like those shown in the previous Section, in which the rotation may become retrograde after crossing a bifurcation. In this appendix, we show that for circular orbits, Darwin's theory for viscous planets and the creep tide theory coincide, and that other Darwinian approaches do not lead to results different from those found using the creep tide theory to calculate the tidal variation of the rotation speed. 

The use of a plugged ad hoc lag instead of a lag calculated from first principles is a simplification introduced by Darwin himself  (Darwin, 1880). The starting point in the theories referred to currently as `Equilibrium Tide theories' is the static tide. The dynamic tide is assumed to be formed by the same harmonics as the static tide, each delayed by a given phase delay, $\varepsilon_k$ (see  Ferraz-Mello, 2019). Besides, each term is assumed to have an amplitude attenuation equal to the cosine of the corresponding phase delay. In Darwin's words: 
\textit{In a frictional fluid, the tide will be reduced in height and altered in phase.} (Darwin, 1880, \S{4}). 

If the weak-friction approximation\footnote{The 'weak-friction approximation', introduced by  Gerstenkorn (1955), assumes that the lags ($\varepsilon_k$) are small, allowing the simplifications $\sin{\varepsilon_k} = \varepsilon_k$ and $\cos{\varepsilon_k}=1$. These apparently harmless modifications introduce essential changes in the theory and are at the origin of several conceptual difficulties (see Efroimsky and Williams, 2009, Sec. 9.4)}
is \underline{not} used, the average contribution of the tides to the planet rotation (see Ferraz-Mello et al. 2008) becomes:
\begin{equation}
    \langle \dot{\Omega} \rangle = 
    - \frac{3k_2GM^2R^5}{4Ca^6}\sin 2\varepsilon_0
    \equiv - A\frac{2\tan \varepsilon_0}
        {1+\tan^2\varepsilon_0}
\label{eq:Darwin}
\end{equation}
in the circular approximation. 

\subsection{Darwin's theory for a viscous planet}

In the case of a viscous planet, the plugged ad hoc lag is 
$$\varepsilon_0=\arctan \tau\nu $$
where $\tau$ is a constant (Darwin, 1880, \S 7), and the tidal variation of the planet's rotational velocity is   
\begin{equation}
    \langle \dot{\Omega} \rangle = 
 - A\frac{2\tau\nu }
        {1+\tau^2\nu^2}.
\label{eq:Darwin}
\end{equation}

If we make $\tau=\frac{1}{\gamma} $, this equation is transformed into the same equation obtained with the creep tide theory in the circular approximation. Therefore, the stationary solutions for the rotation of a planet under the joint action of the gravitational and atmospheric tides are the same as those given in Section \ref{sec:atmosf}.

\subsection{The Maxwell and Andrade rheologies}

\begin{figure}[h]
	\centering
	\includegraphics[width=0.4
    \textwidth]{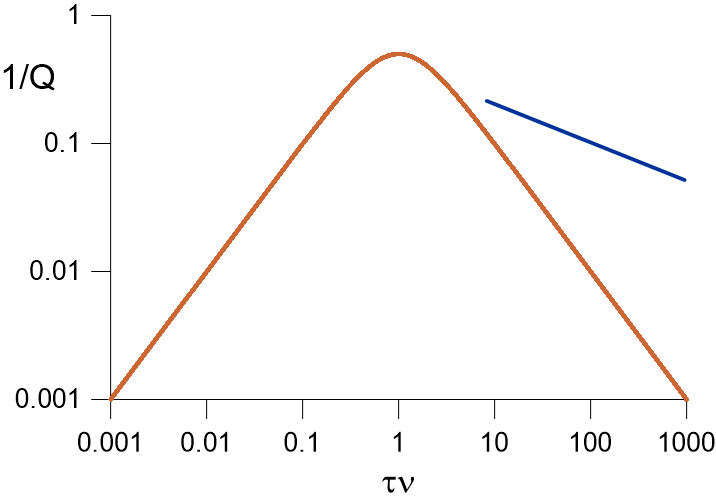}
	\caption{Maxwell rheology: Log-log graphic of the inverse of the Quality Factor $Q$ as a function of $\tau\nu$. The blue line shows the slope of the descending branch in the case of the Andrade rheology with $p=-0.2$.} \label{fig:Q}
\end{figure}

The general equation (\ref{eq:Darwin}) may be written as
\begin{equation}
   \langle \dot\Omega \rangle  =
   -2A\left(\tan\varepsilon_0+\frac{1}{\tan\varepsilon_0} \right)^{-1}
\end{equation}
or, using Darwin's approach for viscous planets,
\begin{equation}
   \langle \dot\Omega \rangle  =
   -2A\left(\tau\nu+\frac{1}{\tau\nu} \right)^{-1}.
\end{equation}
The function 
\begin{equation}
Q = \left(\tau\nu+\frac{1}{\tau\nu} \right)
\end{equation} 
appears frequently in the existing literature on planetary tides and is referred to as the \textit{Quality factor} (see  Ferraz-Mello, 2015). It carries an interpretation related to the energy dissipation in the system and is characteristic of the Maxwell rheology. Fig. \ref{fig:Q} is the well-known log-log plot of $1/Q$ as a function of $\tau\nu$ for the Maxwell rheology. If no logarithmic scales are used, the figure is exactly equal to the curves in the phase diagram shown in Fig. \ref{fig:omega1} (in an arbitrary scale).   

Laboratory studies indicate that solid bodies do not follow the Maxwell rheology when $|\tau\nu|>>1$, but the Andrade rheology (Efroimsky, 2012a; Gevorgyan et al. 2020). 
However, when the Andrade rheology is adopted, the results are comparable to those of Darwin's and creep tide theories with their intrinsic Maxwellian rheology. The only difference lies in the fact that the descending branch of the Andrade rheology will show an elbow and, after it, a slower decay (as shown by the blue line of Fig. \ref{fig:Q}. Consequently, the slow decay of the tidal component $|d\Omega/dt|$ for larger $|\nu|$ (see Fig. \ref{fig:composit}) will be even slower. However, the interval between the maximum and minimum of $|d\Omega/dt|$ will not be affected (since it corresponds to the ascending branch, which is the same in both rheologies). We may conclude that the asynchronous solutions in the case shown in Fig. \ref{fig:composit} are the same in both Andrade and Maxwell rheologies.   

\subsection{The Constant Time Lag (CTL) and Constant Phase Lag (CPL) theories}

Two widely used versions of Darwin's theory are the Constant Time Lag (CTL) and the Constant Phase Lag (CPL) theories (see Greenberg, 2009; Barnes, 2017). 

In the circular approximation of the CTL model, the plugged lag is a quantity proportional to the semi-diurnal frequency:
    $$\varepsilon_0=\tau\nu, $$
where $\tau$ is a constant of proportionality (time lag). The CTL theory results coincide with the Darwin theory for viscous planets as long as $\tau\nu << 1$.
It can be used to study the stationary synchronous solution and its neighborhood, but not elsewhere. For larger values, the two theories diverge from one another. Besides, the results of the CTL show a periodicity induced by the periodic structure of $\tan\tau\nu $, which is an artifact. 

In the CPL theory, the plugged lag $\varepsilon_0$ does not change with the rotation velocity and is discontinuous when $\nu=0$. For this reason, it cannot be used to study the conversion of a prograde planetary rotation into a retrograde one, which is the aim of this investigation (see Efroimsky, 2012b).


\end{document}